# DEGRADATION OF METHYLENE BLUE USING GRAPHENE OXIDE-TIN OXIDE NANOCOMPOSITE AS PHOTOCATALYST

M.Bakhtiar Azim[a*], Intaqer Arafat Tanim[a], Riad Morshed Rezaul[a], Rizwan Tareq[a],
Arafat Hossain Rahul[a], A.S.W Kurny[b] and Fahmida Gulshan[b*]

[a]Department of Materials and Metallurgical Engineering, Faculty of Engineering, Bangladesh University of Engineering and Technology, Dhaka-1000, Bangladesh

## ABSTRACT

Environmental contamination and human exposure to dyes have dramatically increased over the past decades because of their increasing use in such industries as textiles, paper, plastics, tannery and paints. These dyes can cause deterioration in water quality by imparting color to the water and inducing the photosynthetic activity of aquatic organisms by hindering light penetration. Moreover, some of the dyes are considered carcinogenic and mutagenic for human health. Therefore, efficient treatment and removal of dyes from wastewater have attracted considerable attention in recent years. Photocatalysis, due to its mild reaction condition, high degradation, broad applied area and facile manipulation, is a promising method of solving environmental pollution problems. In this paper, we report the synthesis of graphene oxide-tin oxide (GO-$SnO_2$) nanocomposite and the effectiveness of this composite in decolorizing Methylene Blue. Tin oxide was prepared by liquid phase co-precipitation method and graphene oxide-tin oxide (GO-$SnO_2$) nanocomposite were prepared by solution mixing method. Tin oxide ($SnO_2$) nanoparticles have been ardently investigated as photocatalyst for water purification and environment decontamination but the photon generated electron and hole pair (EHP) recombination is one of the limiting factors. Graphene oxide-tin oxide (GO-$SnO_2$) nanocomposite is very propitious to overcome this limitation for photocatalytic application. The as-synthesized graphene oxide (GO) and GO-$SnO_2$ nanocomposite were characterized by X-ray Diffraction (XRD), Scanning Electron Microscopy (SEM), Energy Dispersive X-ray spectroscopy (EDX). The GO-$SnO_2$ nanocomposite showed better photocatalytic degradation efficiency for Methylene Blue compared to $SnO_2$ and Graphene oxide.

## 1. INTRODUCTION

With the rapid development of textile industry in recent years, more and more new types of dyes, such as Methylene Blue, have been produced. According to a World Bank report, almost 20% of global industrial water pollution comes from the dyeing and finishing processes of textiles [1]. The discharge of azo dyes, which are stable and carcinogenic, into water bodies are harmful to human health, and cause such illness as cholera, diarrhea, hypertension, precordial pain, dizziness, fever, nausea, vomiting, abdominal pain, bladder irritation, staining of skin [2]. Dyes also affect aquatic life by hindering the photosynthesis process of aquatic plants, eutrophication, and perturbation [3, 4]. Numerous techniques, such as activated carbon adsorption (physical method), chlorination (chemical method), and aerobic biodegradation (biochemical method) [4] have been applied to treat textile wastewater. However, further treatments are needed, which create such secondary pollution in the environment, as the breakdown of parent cationic dyes to Benzene, $NO_2$, $CO_2$ and $SO_2$ [6]. Advanced oxidation processes (AOPs) are widely applied to mineralize dyes into $CO_2$ and $H_2O$ [7, 8]. AOPs include ozonation, photolysis, and photocatalysis with the aid of oxidants, light, and semiconductors. Photocatalytic degradation is initiated when the photocatalysts absorb photons (UV) to generate electron-hole pairs on the catalyst surface. The positive hole in the valence band ($h_{VB}^+$) will react with water to form hydroxyl radical (*OH), followed by the oxidization of pollutants to $CO_2$ and $H_2O$ [9].

Methylene Blue (MB), also known as Basic Blue 9, is an azo dye (Table 1). MB is widely used in textile industries for dye processing, and upto 50% of the dyes consumed in textile industries are azo dyes [9-11]. In the past few years, several catalysts such as $TiO_2$ [6], $BiFeO_3$ [5], ZnS [13] and ZnO [9] have been used to degrade MB and the results are summarized in (Table 2).

**Table 1. Properties of Methylene Blue (MB).**

| Properties | Cationic Azo Dye |
|---|---|
| Synonym name | Basic Blue 9 |
| Molecular formula | $C_{16}H_{18}ClN_3S$ |
| Molecular weight | 319.851 g/mol |
| Absorbance wavelength($\lambda_{max}$) | 664 nm |
| Molecular structure | (structure of methylene blue) |

**Table 2. The photocatalytic degradation of MB using several catalysts.**

| Authors/Year | Catalysts | Degradation efficiency (%) | Conditions | References |
|---|---|---|---|---|
| Soltani *et al.* 2014 | $BiFeO_3$ | 100% MB | Time: 80 min; Catalyst loading: 0.5 g/L$^{-1}$; Irradiation: Natural Sunlight; pH 2.5 | [5] |
| Dariani *et al.* 2016 | $TiO_2$ | 100% MB | Time: 2 hr; Catalyst loading: 0.5 g/L$^{-1}$; Irradiation: UV light; pH 2.5 | [6] |

Graphene oxide (GO) has more oxygen functional groups than reduced graphene oxide (rGO). GO has a surface area of 736.6 m$^2$/g [30] compared to 400 m$^2$/g [31] for graphite. Numerous chemical, thermal, microwave and microbial/bacterial methods have been used in the synthesis of GO [15]. Chemical exfoliation is preferable due to its large-scale production and low cost. Chemical exfoliation involves three steps, oxidation of graphite powder, dispersion of graphite oxide (GTO) to graphene oxide (GO) and GTO exfoliation by ultrasonication to produce graphene oxide (GO) [16]. GO, with its unique electronic properties, large surface area and high transparency, contributes to facile charge separation and adsorptivity in its structure. As a potential photocatalytic material, GO-SnO$_2$ has been used in the decolorization of Methylene Blue [20] and Rhodamine B [20].

SnO$_2$ is a capable candidate to photo catalyze and complete oxidative mineralization of MB. Chemical contamination of water streams has become crucial issue of human life. Wastewater treatment plays an important role in reducing the toxic elements in wastewater. Heterogeneous photocatalysis can be applied to remove contaminants existing in wastewater effluent; catalysis under natural sunlight irradiation is called photocatalysis. Heterogeneous photocatalysis includes such reactions as organic synthesis, water splitting, photo-reduction, hydrogen transfer and metal deposition, disinfection, water treatment, removal of gaseous pollutants etc. It has become an increasingly viable technology in environmental decontamination. Photocatalytic oxidation of such organic compounds are derived by semiconductor materials like TiO$_2$, ZnO, CdS and CuO. These are now-a-days widely used in the environment as photocatalysts. They have band-gap of 3.6 eV, when other photocatalysts have higher band-gap. They are used extensively due to their low-cost, non-toxicity, high activity, large chemical stability, very low aqueous solubility and environmental friendly characteristics. Tin-assisted photocatalytic oxidation is an alternative method for purification of air and water streams. Also, in water splitting, the driving force for electrons is provided by energy of light. When SnO$_2$ is exposed to light, photocatalytic reaction is initiated. GO-SnO$_2$ nanocomposite is very promising to overcome the limitation for photocatalytic application. GO with its large surface area, unique electronic properties and high transparency, contributes to spatial charge separation and adsorptivity in this hybrid structure.

In this investigation, we report a facile method to prepare GO-SnO$_2$ nanocomposite. It was synthesized via solution mixing method. The photocatalytic performances of the prepared GO, SnO$_2$ and GO-SnO$_2$ nanocomposite were evaluated in the degradation of a model organic dye, methylene blue (MB), in aqueous solution under natural sunlight. To best of our knowledge, detailed investigations on catalyst loading, initial dye concentration and initial solution pH are still lacking. This study aims to determine the optimum experimental conditions for the best photo-decolorization performance.

## 2. EXPERIMENTAL SECTION

### 2.1. CHEMICALS AND MATERIALS

Graphite fine powder, Sodium Nitrate, Tin (II) Chloride Di-hydrate (98%), Potassium permanganate (99%), Hydrochloric acid (37%), Hydrogen Peroxide (30%), and Ammonia solution (25%) were purchased from Sigma-Aldrich (Steinheim, Germany). Sulfuric acid (98%) was obtained from Merck (Darmstadt, Germany). The chemicals were used without further purifications. Methylene Blue (MB) powder from Sigma-Aldrich (Steinheim, Germany) was used as the model compound in this study. Deionized water was used throughout the experiments.

### 2.2. SYNTHESIS OF GRAPHENE OXIDE (GO)

Graphene oxide was produced through the modified Hummers' method by oxidizing the graphite powder. In a typical synthesis, 5g of graphite powder and 2.5g $NaNO_3$ were mixed with 115 ml $H_2SO_4$ (conc. 98%). 15g of $KMnO_4$ was slowly added and stirred in an ice-bath for 1 h below 20°C. The mixture was heated to 35°C and was constantly stirred for 2 hrs. The beaker was placed on an oil bath, heated to and maintained at a temperature 95°C~ 98°C, for 15 minutes. 250 ml DI water was added slowly under constant stirring. The mixture was cooled to room temperature. The beaker was then placed on the oil bath for additional 60 minutes at a constant temperature 60°C. 150 ml DI water was added under constant stirring. Finally, 50 ml (30%) $H_2O_2$ was added in drops and stirred for 2 hrs. Washing, filtration and centrifugation were done until the removal of $Cl^-$ ions by using DI water. Finally, the resulting precipitate was dried at 70°C for 24 hrs in an oven giving thin sheets which was Graphite Oxide (GTO). Graphite Oxide was made into fine powder form by grinding and then GTO powder was finely dispersed in DI Water. At last, ultrasonication was done for the complete exfoliation of GTO to GO.

### 2.3. SYNTHESIS OF TIN OXIDE ($SnO_2$)

$SnO_2$ was produced through the liquid phase co-precipitation method. 2g Stannous Chloride Di-hydrate ($SnCl_2.2H_2O$) was dissolved in 100 ml DI Water. After complete dissolution, ammonia solution (25%) was added in drops to the above solution under stirring. The resulting gel type precipitate was filtered and dried at 80°C for 24 hours to remove water molecules. Finally, tin oxide nanopowders were formed through calcination at 550°C for 4-6 hours.

### 2.4. SYNTHESIS OF GRAPHENE OXIDE-TIN OXIDE NANOCOMPOSITE (GO-$SnO_2$)

GO-$SnO_2$ was prepared through the solution mixing method. At first, 2.6 gm $SnCl_2.2H_2O$ was added to HCl (37%, 1.4 ml) and stirred for 1 hour. Then, 260 mg GO was dispersed in 200 ml DI water by using ultrasonication for 45 minutes. Both solutions were mixed and stirred rigorously for 15 minutes. Again ultrasonication was done for 15 minutes. After that, washing, filtration and centrifugation were done with DI water till a neutral pH was obtained. The sediment was collected and dried at 80°C for 24 hours. Finally, thin sheet of GO-$SnO_2$ nanocomposite was collected and grinded into GO-$SnO_2$ nano powder.

### 2.5. CHARACTERIZATION

The X-ray diffraction pattern of GO, $SnO_2$ and GO-$SnO_2$ were recorded by a Bruker, D8 Advance diffractometer (Germany). The sample was scanned from 5° to 80° using Cu Kα radiation source (λ = 1.5406 A°) at 40 kV and 30 mA with a scanning speed of $0.01°s^{-1}$. The surface morphology of GO, $SnO_2$ and GO-$SnO_2$ was observed by HITACHI (Tokyo, Japan) TM3030 table-top scanning electron microscope (SEM). The decolorization percentage of MB was determined by using an ultraviolet-visible spectrophotometer (Agilent Cary-60, Santa Clara, CA, USA) at $\lambda_{max}$ = 664 nm a wavelength region between 400 and 800 nm. DI water was used as a reference material.

### 2.6. PHOTOCATALYTIC REACTION

Photocatalytic experiments were carried out by photodegrading MB using UV-Vis spectroscopy. A 96 W UV-A lamp was used as the irradiation source. The solution of MB (pH~7) without GO-$SnO_2$ was left in a dark place for 60 min.

Then, the dye solution was exposed to natural sunlight irradiation and there was no decrease in the concentration of dye. In a typical experiment, 7.5 mg of GO-$SnO_2$ was added into a 50 mL 0.05 mM MB solution. Before illumination, the suspensions were continuously stirred in dark for 60 min to reach an adsorption-desorption equilibrium between the photocatalyst and MB. The suspensions were then exposed to natural sunlight irradiation for another 60 mints and samples were taken at regular time intervals (0 min, 10 min, 15 min, 30 min, 45 min and 60 min) and filtered to remove the GO-$SnO_2$. Irradiation was carried out in a volumetric flask. The same procedure was repeated for both GO and $SnO_2$. Where required, the initial pH of solution (pH~7) was adjusted by small amount of 0.1 M NaOH and 0.1 M HCl. Photodegradation was also observed for 0.05 mM MB solution using only $SnO_2$ and GO. The decolorization efficiency of MB was determined by using the equation shown below:

$$\textbf{Photodegradation efficiency (\%)} = [(C_0 - C_t) / C_0] \times 100\% = [(A_0 - A_t) / A_0] \times 100\% \quad (1)$$

where $C_0$ is the initial concentration of MB, $C_t$ is the concentration of MB at time, t and $A_0$ is the initial absorbance of MB, $A_t$ is the absorbance of MB at time, t.

## 3. RESULTS AND DISCUSSIONS

### 3.1. CHARACTERIZATION

The powder X-ray diffraction pattern of GO shows a broadened diffraction peak (Fig. 1(b)) at around $2\theta \approx 10.46°$, which corresponds to the (002) reflection of stacked GO sheets. XRD patterns of $SnO_2$ nanoparticles (Fig. 1(c)) shows the diffraction peaks of (110), (101), (111), (211), (220), (002), (310), (112), (301), (202) and (321) at $2\theta$ of 26.8°, 33.9°, 37.9°, 51.8°, 54.8°, 57.7°, 61.8°, 64.8°, 66.0°, 71.2° and 78.6° respectively which matches well with JCPDS card# 41-1445. The average crystallite size of $SnO_2$ was determined using "Scherrer Formula" and its value was found to be ~8 nm.

$$B = (k\lambda)/(\beta cos\theta_\beta) \quad (2)$$

where, B is the average crystallite size (nm), k is the factor of 0.9, λ is the wavelength of radiation source used is 1.5406 A°, β is the full width of half maximum peak (FWHM), $\theta_\beta$ is the angle at maximum peak. The XRD pattern of GO-$SnO_2$ nanocomposite (Fig. 1(d)) shows only $SnO_2$ peaks without GO peak in XRD pattern of GO-$SnO_2$ nanocomposite has been observed, but the intensity of the peaks are reduced and broaden when compared with XRD pattern of individual GO and $SnO_2$.

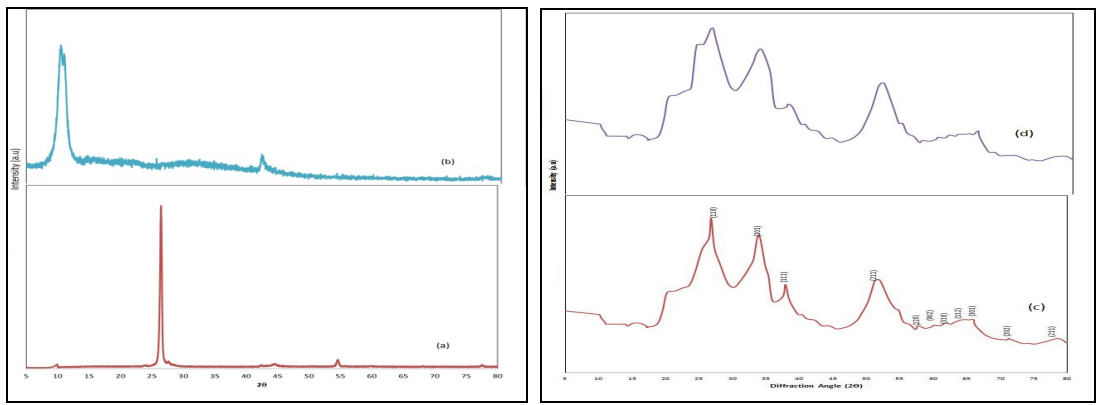

FIG. 1: XRD PATTERN OF (a) GRAPHITE POWDER, (b) GRAPHENE OXIDE (GO), (c) TIN OXIDE ($SnO_2$) and (d) GO-$SnO_2$.

SEM images of GO-$SnO_2$ structure are shown in (Fig. 2). SEM images of GO-$SnO_2$ shows presence of $SnO_2$ nanoparticles in the crumbled sheet of GO layers.

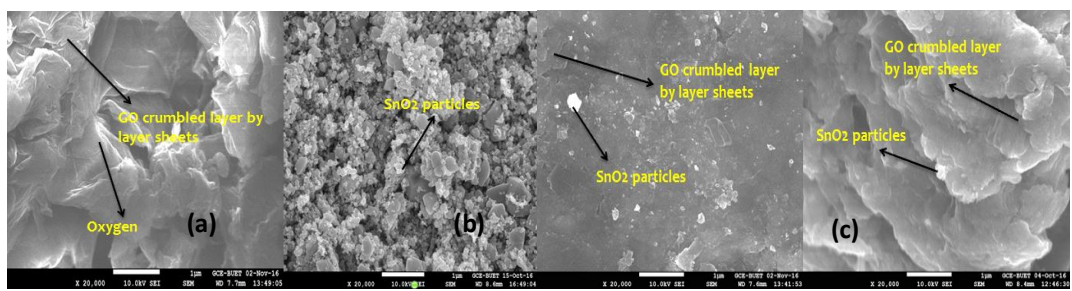

FIG. 2: SEM IMAGE OF (a) GRAPHENE OXIDE (GO), (b) $SnO_2$ NANOPARTICLES and (c) GO-$SnO_2$ NANOCOMPOSITE

EDX results indicating successful incorporation of oxygen by modified Hummers method in graphite layers and formation of oxygen based functional groups in GO with ratio of C/O=4.58 (Fig. 3(b)). It also shows successful synthesis of $SnO_2$ nanoparticles (Fig. 3(c)). It gives clear indication of incorporation of $SnO_2$ nanoparticles in GO crumbled layer by layer sheets (Fig. 3(d)).

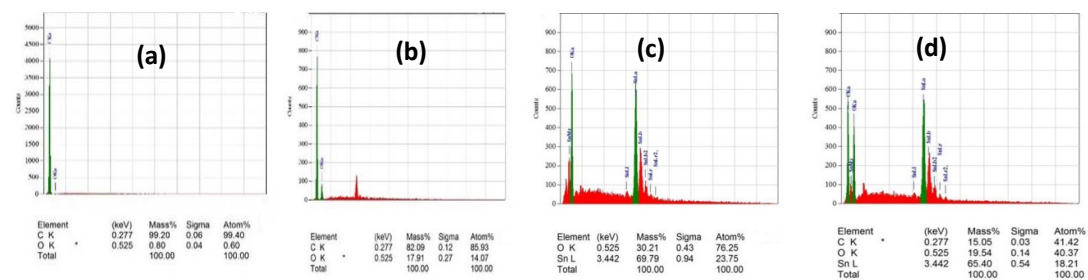

FIG. 3: EDX SPECTRA OF (a) GRAPHITE POWDER, (b) GRAPHENE OXIDE (GO), (c) $SnO_2$ NANOPARTICLES and (d) GO-$SnO_2$ NANOCOMPOSITE

3.2. PHOTOCATALYTIC ACTIVITY OF GO-$SnO_2$

UV-Vis was used to measure absorbance of the dye solution at regular time intervals using GO-$SnO_2$. Controlled experiments were also carried out to confirm that the degradation of MB by UV-Vis for visible range. Experiments were repeated for only GO and for only $SnO_2$. Under natural sunlight irradiation GO-$SnO_2$ nanocomposite showed 89.4% photodegradation efficiency after 60 min whereas only GO and only $SnO_2$ showed 68.68% and 62.81% respectively (Fig. 4).

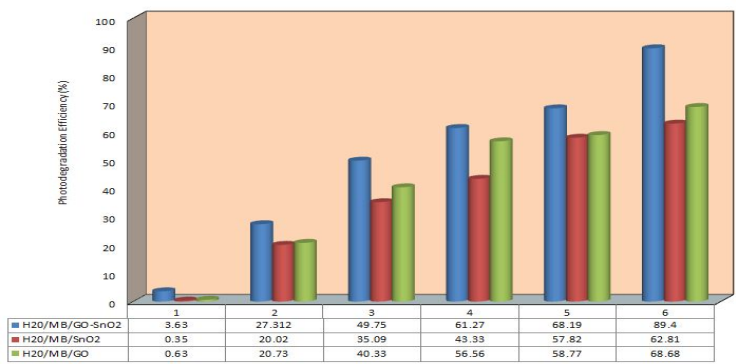

FIG. 4: PHOTO DEGRADATION EFFICIENCY OF THE $H_2O$/MB/GO, $H_2O$/MB/$SnO_2$ AND $H_2O$/MB/GO-$SnO_2$.

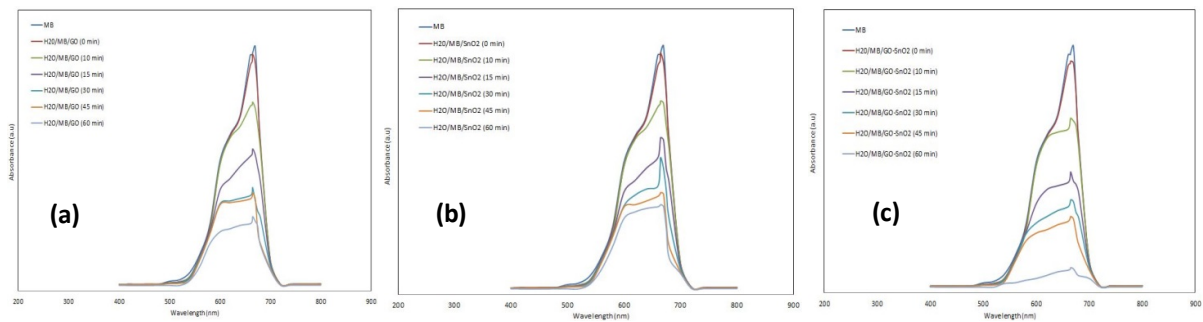

FIG. 5: TIME-DEPENDENT ABSORPTION SPECTRA OF MB SOLUTION DURING NATURAL SUNLIGHT IRRADIATION IN THE PRESENCE OF (a) $H_2O/MB/GO$, (b) $H_2O/MB/SnO_2$ AND (c) $H_2O/MB/GO-SnO_2$.

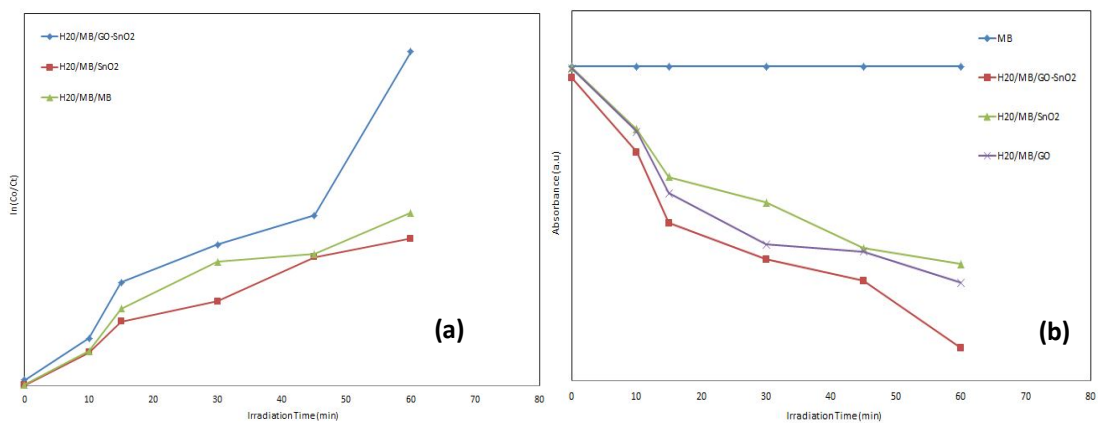

FIG. 6: (a) $\ln(C_o/C_t)$ VERSUS IRRADIATION TIME AND (b) ABSORBANCE VERSUS IRRADIATION TIME CURVES ILLUSTRATING MB DEGRADATION by $H_2O/MB/GO$, $H_2O/MB/SnO_2$ AND $H_2O/MB/GO-SnO_2$.

It is clear from Fig. 6(a) and Fig. 6(b) that, GO-$SnO_2$ nanocomposite as expected showed highest photocatalytic activity compared to that of $SnO_2$ and GO. GO-$SnO_2$ nanocomposite is an active photocatalyst for MB degradation which takes 60 min for almost total (~89%) degradation. Photocatalytic activity of GO-$SnO_2$ can be studied for further times.

Semiconductor-based photocatalyst relies on the active absorption of a photon from a semiconductor to create an electron-hole pair and this process depends on the band gap of the material. Then the excited electron has to be separated from the hole to avoid recombination. The electron can also be used to reduce chemicals in the environment by generating radical species such as hydroxyl radicals; which can initiate, for example, degradation reactions. GO is an insulating and p-type material. Due to disturbance of its $sp^2$ bonding structure and having high surface area (~740 $m^2/g$); it can be used as a photocatalyst by doping with metal oxide like $SnO_2$. As a result, it reduces the band gap of GO, creating electron hole-pair and prevents recombination.

## 4. CONCLUSION

Degradation of Methylene Blue under natural sunlight with GO-$SnO_2$ nanocomposite as a photocatalyst takes around 60 min for ~89% degradation compared to ~69% for GO and ~63% for $SnO_2$. Photodegradation was investigated for 50 ml 0.05 mM MB solution and 7.5 mg of GO, $SnO_2$ and GO-$SnO_2$ each. It can be studied for further times and optimum amount for total Methylene Blue degradation under natural sunlight.

## 5. ACKNOWLEDGEMENT

This work was financially supported by Bangladesh University of Engineering and Technology (BUET). The authors are also grateful to the Department of Glass and Ceramics Engineering (GCE, BUET), Department of Chemistry, BUET and Department of Materials and Metallurgical Engineering (MME, BUET).